\documentclass[12pt]{article}
\usepackage{amssymb,amsmath,xcolor,graphicx,xspace,colortbl, rotating} 
\usepackage{amsfonts}  
\usepackage{amsmath}  
\usepackage{amssymb}  
\usepackage{epsfig,graphicx}  
\usepackage[authoryear]{natbib}
\usepackage{latexsym}  
\usepackage{rotating}  
\usepackage{setspace}  
\usepackage{mathtools}

\usepackage{hyperref}

\usepackage{enumitem}
\usepackage{wrapfig}

\usepackage{amsthm}

\usepackage{thmtools,thm-restate}

\newtheorem{assumption}{Assumption}

\graphicspath{{MMLV_graphics/}{MMLV_tcache/}{MMLV_gcache/}}
\DeclareGraphicsExtensions{.pdf,.svg,.eps,.ps,.png,.jpg,.jpeg}
\usepackage {bm}
\usepackage {tabulary}
\newtheorem {theorem}{Theorem}

\newtheorem {corollary}{Corollary}

\newtheorem {definition}{Definition}

\newtheorem {lemma}{Lemma}

\newtheorem {proposition}{Proposition}

\begin{document}

\title{Identification of Multivariate Measurement Error Models}

\author{
Yingyao\ Hu\protect\footnote{First version: December 2, 2025. Contact information: Department of Economics, Johns Hopkins University, 3400 N. Charles Street, Baltimore, MD 21218. Email: yhu@jhu.edu.  } 
}
\date{\today}
\maketitle
\begin {abstract}

This paper develops new identification results for multidimensional continuous measurement-error models where all observed measurements are contaminated by potentially correlated errors and none provides an injective mapping of the latent distribution. Using third-order cross-moments, the paper constructs a three-way tensor whose unique decomposition, guaranteed by Kruskal’s theorem, identifies the factor loading matrices. Starting with a linear structure, the paper recovers the full distribution of latent factors by constructing suitable measurements and applying scalar or multivariate versions of Kotlarski’s identity. As a result, the joint distribution of the latent vector and measurement errors is fully identified without requiring injective measurements, showing that multivariate latent structure can be recovered in broader settings than previously believed. Under injectivity, the paper also provides user-friendly testable conditions for identification. Finally, this paper provides general identification results for nonlinear models using a newly-defined generalized Kruskal rank - signal rank - of intergral operators. These results have wide applicability in empirical work involving noisy or indirect measurements, including factor models, survey data with reporting errors, mismeasured regressors in econometrics, and multidimensional latent-trait models in psychology and marketing, potentially enabling more robust estimation and interpretation when clean measurements are unavailable.


\bigskip 


\end {abstract}

\setcounter{page}{0}
\thispagestyle{empty}

\newpage
\setstretch {1.15}

\section{Introduction}


Multivariate continuous latent variables arise in numerous empirical settings in economics, psychology, marketing, epidemiology, and the social sciences. Examples include multidimensional skills, cognitive factors, latent preferences, health indices, productivity components, and risk attitudes. In practice, such latent constructs are rarely observed directly; instead, researchers rely on multiple imperfect measurements that contain potentially correlated forms of noise. The resulting measurement-error problem is especially severe when the latent variable is multidimensional: each measurement typically captures only a low-dimensional projection of the latent vector, and the noise may exhibit dependence or heterogeneity across measurement channels. As a result, classical approaches to continuous measurement error, which often rely on injectivity, offer limited guidance.

This paper develops an identification strategy tailored specifically to multivariate continuous latent variables measured with noise. The key innovation is to combine tools from multilinear algebra, specifically the uniqueness properties of so-called CP tensor decompositions, with the multivariate extension of Kotlarski's identity, a powerful deconvolution result based on characteristic functions. The starting point is the observation that third-order cross-moments of three separate measurements form a three-way moment tensor whose CP decomposition is governed by the latent factor loading matrices. By invoking Kruskal's theorem \citep{Kruskal1977}, I show that these loadings are generically identifiable even when each measurement matrix is rank-deficient and---even more surprisingly---when the stack of all measurement matrices is non-injective. This greatly relaxes the standard requirements for identifying multivariate continuous latent variables, which typically demand injectivity of the measurement operator.

Once the factor loading matrices are recovered through tensor methods, the paper turns to the problem of identifying the entire joint distribution of the continuous latent vector. A crucial step of the identification is that the Kruskal rank condition leads to two conditionally independent noisy measurements of each coordinate of the latent vector, derived from suitable linear projections of the observed measurements. These constructed pairs satisfy the assumptions of Kotlarski's identity, enabling closed-form identification of each coordinate's marginal distribution. For the case of dependent continuous latent factors, a weaker version of the full rank condition allows us to produce two vector-valued measurements whose errors are independent of the latent vector, thereby permitting the use of the multivariate Kotlarski identity. This yields identification of the full joint distribution of the continuous latent vector without assuming independence across its components.

A major contribution of this paper lies in demonstrating that multivariate continuous latent variables can be identified nonparametrically from three non-injective measurements by combining tensor uniqueness with characteristic-function deconvolution. It can be considered as a small step towards the extension of the Kruskal's theorem to continuous settings. As \cite{allman2009identifiability} showed, the Kruskal's theorem can identify distribution of discrete latent variables. Extending that to continuous variables would require the extension of the Kruskal's theorem to continuous setting, which, to the best of my knowledge, is still an open question. The theorem in \cite{hu_schennach}  can be considered as such an extension to the continuous case under injectivity because a full rank condition in the discrete case naturally extends to the injectivity condition in the continuous case. Furthermore, another important contribution of this paper is to provide a generalized Kruskal rank - signal rank - for integral operators, and show that identification can hold for general nonlinear models under a signal rank condition similar to the well-known Kruskal rank condition. The newly-defined rank is consistent with the existing Kruskal rank for matrices and the injectivity of integral operators, and can be used to describe how informative a measurement is in the continuous multivariate case. Based on the signal rank condition, I provide identification of nonlinear models under high-level assumptions. In summary, this paper provides a constructive identification solution to a linear multivariate measurement error model, and builds a foundation for a better solution to the identification of general nonlinear measurement error models.

This work is closely related to several strands of literature. Nonparametric identification of measurement-error models with continuous latent variables has been advanced by e.g., \cite{kotlarski1965pairs},  and \cite{hu_schennach}. Tensor decomposition methods have been widely used to handle discrete latent variables in psychometrics, e.g., \cite{CarrollChang1970,Harshman1970}, signal processing, e.g., \cite{Sidiropoulos2000}, and in econometrics and statistics, e.g., \cite{Hu2008discrete}, \cite{allman2009identifiability} s,\cite{BonhommeJochmansRobin2016}.  This paper contributes a novel bridge between these literatures by demonstrating how tensor uniqueness can be used to identify both the measurement system and the full distribution of a multivariate continuous latent variable without injectivity or with less injectivity restrictions.

The limitation of this paper lies in the high-level assumptions in the nonlinear cases. The linear specification of the multivariate measurement structure may be rich enough for most empirical applications, but not complicated enough for theoretical explorations. Relaxaion of the linearity and additivity requires new technical tools. For that purpose, this paper provides a promising starting point towards continuous or operator-valued generalizations of Kruskal's theorem. Such continuous tensor identifiability results would naturally complement the multivariate Kotlarski identity and could relax the need for discretized third-order moments. With the results and the new concept in this paper, we are a small step closer toward a unified theory of identification for high-dimensional continuous latent-variable models in the presence of complex, nonclassical measurement error.

This paper is organized as follows: Section 2 specifies the linear model; Section 3 shows the identification of the factor loadings using the Kruskal's theorem; Section 4 presents the identification of continuous distributions without injectivity, but also provides a user friendly identification results under injectivity; Section 5 uses a newly-defned signal rank to present the identification of nonlinear cases. Section 6 concludes. All the proofs are in the Appendix.

\section{A 3-measurement model and injectivity}

We are interested in a latent structure described by a vector of latent variables 
$$X^*=\begin{pmatrix}
  X^*_1 \\
  ... \\
  X^*_L
\end{pmatrix}\in \mathbb{R}^L$$ 
Instead of $X^*$, we observe in a sample of three multi-dimensional measurements, for $i=1,2,3$
$$X^i = \begin{pmatrix}
  X_{1}^{i} \\
  ... \\
  X_{K_i}^{i}
\end{pmatrix} \in \mathbb{R}^{K_i}$$
The relationship between the latent vector $X^*$ and its measurements can be described by the integral operator $L_{X^i|X^*} : \mathcal{L}^2\left( \mathcal{R}^{L }\right)
\rightarrow \mathcal{L}^2\left( \mathcal{R}^{K^i}\right)$
\begin{eqnarray}
p_{X^i} &= &\int p_{X^i|X^*}p_{X^*}dX^* \nonumber \\
&\equiv& L_{X^i|X^*} p_{X^*}  
\end{eqnarray}
where $p_{X^*}$ stands for the probability density function of $X^*$. It is known that nonparametric identification of the distribution of the latent variable $X^*$ requires that the integral operator $L_{X^i|X^*}$ is injective. This paper presents the identification of the joint distribution of $X^*$ and $X=\left ((X^1)^T,(X^2)^T,(X^3)^T \right)^T$ without such injectivity or with fewer injectivity restrictions than in the existing literature.

We start with a linear specification of the relationship between the latent vector $X^*$ and the three measurements as follows: 
\begin{eqnarray} \label{equation model}
X^1 &=& M^1 X^* + \varepsilon^1   \\
X^2 &=& M^2 X^* +\varepsilon^2 \nonumber \\
X^3 &=& M^3 X^* +\varepsilon^3  \nonumber
\end{eqnarray}
where $M^i$ are factor loading matrices.
In the case where the dimension of $X^i$ is smaller than that of $X^*$, integral operator $L_{X^i|X^*}$ can't be injective if $p(X^i|X^*)$ is continuous in all its arguments. In particular, when the factor loading matrix $M^i$ doesn't have a full column rank, the injectivity of $L_{X^i|X^*}$ fails. Furthermore, the injectivity may still fail even if we combine the three measurements. For example, we consider 
$L_{X|X^*} :\mathcal{L}^2\left( \mathcal{R}^{L }\right)\rightarrow \mathcal{L}^2\left( \mathcal{R}^{K^1+K^2+K^3}\right) $ 
\begin{eqnarray}
p_{X} &= &\int p_{X|X^*}p_{X^*}dX^* \nonumber \\
&\equiv& L_{X|X^*} p_{X^*}  
\end{eqnarray}
A necessary condition for the injectivity of $L_{X|X^*}$ is that matrix $M$ has a full column rank, where $$ M=\begin{pmatrix}
    M^1 \\M^2 \\M^3
\end{pmatrix}.$$
However, it is possible that $M^1=M^2=M^3$ with $\mathrm{Rank} [M^i] <L$ so that $\mathrm{Rank} [M] <L$. That means the rank of the stacked matrix is the same as that of $M^i$ and the integral operator $L_{X|X^*}$ is still not injective. Such a special case is important for the situation where $X^i$ are repeated measurements, which may have the same factor loading matrix $M^i$. In such a case one can't generate an injective measurement by combining the measurements.

This paper provides a set of sufficient conditions under which the model is fully identified, even when there are no injective measurements, single or combined. We show identification in two steps. First, we provide sufficient conditions to identify the factor loadings. Then, we provide stronger assumptions to show identification of distribution of latent variables without injectivity. In addition, we also show what we can achieve with injectivity under some testable assumptions.

\section{Identification of factor loadings}

We first identify the factor loadings under assumptions as follows:

\begin{assumption} \label{assumption cond mean independence} The following conditions hold:
\begin{enumerate}
\item Each of three error vectors is conditional mean independent of other latent vectors, i.e., for $i \neq j \neq k$
\begin{equation}
    E[\varepsilon^i|\varepsilon^j,\varepsilon^k,X^*]=0.
\end{equation}
The elements in each vector $\varepsilon^i$ are correlated arbitrarily.
\item \label{condition for joint 3rd moments}The latent vector $X^*$ satisfy: 
\begin{eqnarray}
E[X^*]&=&0 \nonumber \\
E[X^*_{-i}(X^*_{-i})^T|X^*_i]&=&E[X^*_{-i} (X^*_{-i})^T] \nonumber\\
E [(X^*_i)^3 ] &\neq &0
\end{eqnarray}
where $X^*_{-i}$ denotes the vector $X^*$ with its $i$-th component removed.
\end{enumerate}
\end{assumption}

In order to identify the factor loadings, we only need the measurement error vectors to be conditional mean independent of each other and the common factors $X^*$. The condition that unconditional mean is equal to zero holds without loss of generality in this linear setting. Notice that the measurement error in different dimension in one multivariate measurement can be correlated with each other arbitrarily. Therefore, we can't split a multivariate measurement to generate more measurements satisfying the same conditions.

Assumption \ref{assumption cond mean independence}.\ref{condition for joint 3rd moments} impose restrictions on the first three moments of the latent vector $X^*$. Assumption \ref{assumption cond mean independence}.\ref{condition for joint 3rd moments} guarantees that for $i\neq j$ and $i\neq k$
\begin{equation}
    E [X^*_i \cdot X^*_j \cdot X^*_k ] = 0.
\end{equation}
which allows us to focus on nonzero third moments of $X^*$ for further identification of factor loadings.

A key observation is that Assumption \ref{assumption cond mean independence} implies 
\begin{eqnarray} 
E (X_i^{1} \times X_u^2 \times X_v^3) 
&=&  \sum_{l=1}^{l=L} m^1_{i,l} \times m^2_{u,l} \times m^3_{v,l} \times   E [(X^*_l)^3]
\end{eqnarray}
which presents a decomposition of a 3-way tensor on the left hand side. The Kruskal's Theorem \citep{Kruskal1977} implies that the elements on the right hand side, i.e., $M^1$, $M^2$, $M^3$, are unique up to permutation and scaling if the Kruskal ranks of  $M^1$, $M^2$, $M^3$ satisfy 
\begin{equation} \label{equation: k rank}
    \kappa(M^1)+\kappa(M^2)+\kappa(M^3) \geq 2L+2
\end{equation}
where $\kappa(M^i)$ defined as
 $$\kappa(M^i)=\max \big\{ k : \text{every set of } k \text{ columns of } M^i
\text{ is linearly independent} \big\}.$$

The Kruskal rank of $M^i$ describes how much information the measurement $X^i$ has even when $M^i$ is not injective or doesn't have a full column rank.  Kruskal’s Theorem implies that we can identify the factor loadings from the decomposition of the 3-way tensor. For example, identification is possible with $K_i=4$ and $L=5$ when $M^i$ has a full column rank $\kappa(M^i)=K^i=4$. That means three 4-dimensional measurements may recover a 5-dimensional latent vector. In summary, we have

\begin{lemma} \label{Lemma loading}
        Under Assumption \ref{assumption cond mean independence}, factor loading matrices $M^1$, $M^2$, $M^3$ in Equation \ref{equation model} are unique up to permutation and scaling if $$\kappa(M^1)+\kappa(M^2)+\kappa(M^3) \geq 2L+2.$$
\end{lemma}
\textbf{Proof}: See the Appendix.

\section{Identification of distributions without injectivity}

In this section, we show identification of distributions of latent variables without injectivity under assumptions as follows:

\begin{assumption} \label{assumption bench0}
Probability function $p(X^*,\varepsilon^1,\varepsilon^2,\varepsilon^3)$ is  continuous and satisfies
\begin{enumerate} 
\item $p(\varepsilon^1,\varepsilon^2,\varepsilon^3|X^*)  = p(\varepsilon^1) \times  p(\varepsilon^2) \times  p(\varepsilon^3)$.
\item $p(X^*)=p(X^*_1) \times ...\times p(X^*_L).$ 
\item $\phi_{X^i}\neq 0$ on $\mathbb{R}^{K_i}$ for $i=1,2,3$, where $\phi_{X^i}$ is the characteristic function of $X^i$.
\end{enumerate}
\end{assumption}
This assumption suggests that the latent variable $X^*$ and the measurement errors $\varepsilon^1$, $\varepsilon^2$, and $\varepsilon^3$ are jointly independent and that the multi-dimensional latent vector contains jointly independent factors. Such a independence allows us to focus on these factors one by one. First, we find two scalar measurements of one factor. Then, we show that the Kruskal rank condition in Equation (\ref{equation: k rank}) guarantees that the two measurements satisfy the conditions to use the Kotlarski's identity. 

Given the definition of the Kruskal rank, this identification procedure applies to all the latent factors with any permutation of the columns or $X^*_l$ for $l=1,...,L$. The distributions of measurement errors can then be identified by deconvolution under that Assumption \ref{assumption bench0}.3, which implies $\phi_{\varepsilon^i} \neq 0$ due to $\phi_{X^i}=\phi_{\varepsilon^i}\phi_{M^i X^*} $. As shown in \cite{HuShiu2022}, this condition is testable. Then, distributions of $X^*$, $\varepsilon^1$, $\varepsilon^2$ are nonparametrically identified with a closed-form solution as follows:
\begin{eqnarray} 
\phi _{X^*} &= &\phi _{X^*_1}\times ...\times\phi _{X^*_L} \\
\phi_{\varepsilon^i} &=& \frac{\phi_{X^i}}{\phi_{M^i X^*} }
\end{eqnarray}

In summary, we have
\begin{theorem} \label{theorem identification distr}
    Suppose that Assumptions \ref{assumption cond mean independence} and \ref{assumption bench0} hold. Then, factor loading matrices $M^i$ for $i=1,2,3$  are unique up to permutation and scaling in Equation \ref{equation model}  and $p(X^i|X^*)$ for $i=1,2,3$ and  $p(X^*)$  are unique in 
     \begin{equation}
          p(X^1,X^2,X^3) = \int p(X^1|X^*)p(X^2|X^*)p(X^3|X^*)p(X^*)dX^*
     \end{equation}
     if 
     $$\kappa(M^1)+\kappa(M^2)+\kappa(M^3) \geq 2L+2.$$     
\end{theorem}

\textbf{Proof}: See the Appendix.

The Kruskal rank condition may be abstract to practitioners. In the case where all the factor loading matrices $M^i$ have a full rank, we have 
\begin{eqnarray} 
\kappa(M^i)&=& \left\{ \begin{matrix}
K_i \text{ if }  K_i < L\\
L \text{ if }  K_i\geq L
\end{matrix} \right.
\end{eqnarray}
That means practitioners can easily detect whether the rank condition holds. In the meantime, we can also interpret the rank condition as how many latent factors the data can identify. We have

\begin{corollary}
    Suppose that Assumptions \ref{assumption cond mean independence} and \ref{assumption bench0} hold and $M^i$ for $i=1,2,3$ have a full rank. Then, the results in Theorem \ref{theorem identification distr} hold if the number of latent factors satisfies  
    \begin{equation}
        L \leq \frac{K_1+K_2+K_3}{2} -1
    \end{equation}
    where $K_i$ is the dimension of $X^i$.
\end{corollary}

\subsection{An illustration}
In this section, we present an illustrative example to provide more details on how the Kruskal rank condition leads to two classical measurements. Suppose each of the three measurements $X^i$ has $K_i=7$ dimensions with $L=8$ latent common factors $X^*$ with the Kruskal ranks equal to $\kappa_i=6$. Therefore, the Kruskal rank condition holds. 

In this case, $X^1$ is a 7-dimentional vector
\begin{eqnarray} 
X^1  
&=& \begin{pmatrix}
m^1_{k,l} 
\end{pmatrix}_{k=1,...,7;l=1,...,8}
X^*  + \varepsilon^{1} \nonumber \\
&\equiv& \begin{pmatrix}
m^1_{.,1} &m^1_{.,2} &... &m^1_{.,8} \end{pmatrix}
X^*  + \varepsilon^{1} \nonumber \\
&\equiv& \begin{pmatrix} M^1_{7 \times 6}&
m^1_{.,7} & m^1_{.,8}
\end{pmatrix}
X^*  +  \varepsilon^{1} 
\end{eqnarray}
where $M^1_{7 \times 6}$ has a full column rank for any permutation of the columns in $M^1$. Therefore, we can find $Q^1$ so that 
\begin{eqnarray} 
W^1=Q^1X^1  
&\equiv& \begin{pmatrix} \begin{pmatrix}I_{6 \times 6} \\0 \end{pmatrix}& 
q^1_{.,7} & q^1_{.,8}
\end{pmatrix}
X^*  + e^{1} 
\end{eqnarray}
Given the same permutation of the columns in $M^1$,  we find a full column rank matrix from the right side as follows:
\begin{eqnarray} 
X^2  
&=& \begin{pmatrix}
m^2_{k,l} 
\end{pmatrix}_{k=1,...,7;l=1,...,8}
X^*  + \varepsilon^{2} \nonumber \\
&\equiv& \begin{pmatrix}
m^2_{.,1} &m^2_{.,2} &... &m^2_{.,8} \end{pmatrix}
X^*  + \varepsilon^{1} \nonumber \\
&\equiv& \begin{pmatrix}
m^2_{.,1} & m^2_{.,2} & M^2_{7 \times 6}
\end{pmatrix}
X^*  + \varepsilon^{2} 
\end{eqnarray}
and with a $Q^2$
\begin{eqnarray} 
W^2=Q^2X^2  
&\equiv& \begin{pmatrix} q^2_{.,1} & q^2_{.,2} &
\begin{pmatrix}
   0\\ I_{6 \times 6} 
\end{pmatrix} 
\end{pmatrix}
X^*  + e^{2} \nonumber \\
&\equiv& \begin{pmatrix} 
q^2_{1,1} & q^2_{1,2} & 0 &0 \\
q^2_{2,1} & q^2_{2,2} & 1 &0 \\
* & * & 0 & I_{5 \times 5}  
\end{pmatrix}
X^*  + e^{2} 
\end{eqnarray}
Notice that $\kappa(M^2)=6$ implies that 
$$\begin{pmatrix} 
q^2_{1,1} \\
q^2_{2,1}
\end{pmatrix} \neq 0.$$ Otherwise, there exist 6 columns, i.e., the 1st one and the last 5, which are linearly dependent so that the Kruskal rank of $M^2$ can't be 6. Without loss of generality, we let $q^2_{1,1}\neq0$
We then consider the first element in each of the two transformed measurements. 
\begin{eqnarray} 
W^1_1 &=& X^*_1 + q^1_{1,7}X^*_7 + q^1_{1,8}X^*_8 +e^1_1   \\
&\equiv& X^*_1 + e_1 \nonumber\\
W^2_1 &=& q^2_{1,1}X^*_1 + q^2_{1,2}X^*_2 + e^{2}_1  \\
&\equiv& \gamma X^*_1 + e_2 \nonumber
\end{eqnarray}
The only common factor in these two scalar measurements is the first latent factor $X^*_1$. Given $\gamma$, we then identified the distribution of $X^*_1$ from the joint distribution of $(W^1_1,W^2_1/\gamma)$ by Kotlarski’s Identity.

In order to identify the distribution of other factors, we may simply swap the columns and then apply the same procedure. For example, for factor $X^*_3$, we just apply the same procedure to
\begin{eqnarray} 
X^1 &=&  \begin{pmatrix}
m^1_{.,3} &m^1_{.,2} &m^1_{.,1} &m^1_{.,4} &...&m^1_{.,8} \end{pmatrix}
X^*  + \varepsilon^{1} \nonumber \\
X^2  
&=& \begin{pmatrix}
m^2_{.,3} &m^2_{.,2} &m^2_{.,1} &m^2_{.,4} &...&m^2_{.,8} \end{pmatrix}
X^*  + \varepsilon^{2} 
\end{eqnarray}

In summary, we may identify 8 latent factors from three 7-dimensional measurements, while the Kruskal rank corresponding to each measurement is 6, i.e., without injectivity.

\subsection{Extention: Identification with correlated factors }
In many empirical applications, especially in the estimation of structural models, it is important to allow the unobserved variables in $X^*$ to be correlated. In fact, we may extend our results to this case with modified assumptions as follows:

\begin{assumption} \label{assumption identification correlated X^*}
Probability function $p(X^*,\varepsilon^1,\varepsilon^2,\varepsilon^3)$ is continuous and satisfies
\begin{enumerate} 
\item $p(\varepsilon^1,\varepsilon^2,\varepsilon^3|X^*)  =  p(\varepsilon^1) \times  p(\varepsilon^2) \times  p(\varepsilon^3)$.
\item $\phi_{X^1} \neq 0$ on $\mathbb{R}^{K_1}$
\end{enumerate}
\end{assumption}

Because the latent factors may be correlated, we are not able to identify the distribution of the factors one by one. Therefore, we impose full rank restrictions on the loading matrix as follows:
\begin{equation} \label{equation full rank}
    \mathrm{Rank}(M^1)=\mathrm{Rank}\begin{pmatrix}
    M^2 \\M^3
\end{pmatrix}=L
\end{equation}
The Kruskal rank codition in Equation (\ref{equation: k rank}) then becomes 
\begin{equation}
    \kappa(M^2)+\kappa(M^3) \geq L+2
\end{equation}
These restrictions in Equation \ref{equation full rank} implies that we may generate two classical measurements of the whole latent vector $X^*$, to which the multivariate Kotlarski's identity applies. That means the distribution of the latent vector is identified with a closed form. Notice that these rank condition and Assumptions \ref{assumption identification correlated X^*}.2 are testable from the data given the first step identification of the factor loadings. 

Finally, We can then identify the distribution of all the latent random variables under additivity and  independence. We have 
\begin{eqnarray} 
\phi_{\varepsilon^i} &=& \frac{\phi_{X^i}}{\phi_{M^i X^*} }
\end{eqnarray}
where the denominsator is nonzero because $\phi_{X^1}=\phi_{\varepsilon^1}\phi_{M^1 X^*}   \neq 0$ implies that $\phi_{X^*} \neq 0$ with a full rank $M^1$.
In summary, we have 

\begin{proposition}
    \label{theorem identification correlated}
    Suppose that Assumptions \ref{assumption cond mean independence} and  Assumption \ref{assumption identification correlated X^*} hold. Then, factor loading matrices $M^i$ for $i=1,2,3$ are unique up to permutation  and scaling in Equation \ref{equation model}  and $p(X^i|X^*)$ for $i=1,2,3$ and  $p(X^*)$ are unique in 
     \begin{equation}
         p(X^1,X^2,X^3) = \int p(X^1|X^*)p(X^2|X^*)p(X^3|X^*)p(X^*)dX^*
     \end{equation}
     if both $M^1$ and $\begin{pmatrix} M^2 \\M^3 \end{pmatrix}$ have a full column rank $L$ and 
     \begin{eqnarray}
         \kappa(M^2)+\kappa(M^3) &\geq& L+2.
     \end{eqnarray}
\end{proposition}

\textbf{Proof}: See the Appendix.

Here we assume that all the density functions are continuous. Therefore, it is not possible to map from a vector to a lower dimension vector without losing information. Technically, there is a one-to-one mapping from $\mathbb{R}^{K}$ to $\mathbb{R}$, but such a mapping can't be continuous for $K>1$, which makes it less useful for practitioners. Also for this reason, the theorem in  \cite{hu_schennach} applies to the current model under three injectivity restrictions, i.e., for $i=1,2,3$, $M^i$ has a full column rank equal to $L$, i.e., 
$\mathrm{Rank}(M^i)=L,$
in order to identify all  the distributions. The advantage of their results is that they can handle general nonlinear measurement functions, such as 
$X^i=g(X^*,\varepsilon^i).$

Nevertheless, Our Theorem \ref{theorem identification distr} achieves identification of all the distributions without injectivity and Proposition \ref{theorem identification correlated} identifies the model under weaker injectivity conditions. In particular, although it has stronger assumptions, Proposition \ref{theorem identification correlated} is more friendly to practitioner because the full rank condition and the nonzero ch.f. condition are both testable and easy to interpret.

\section{Nonlinear cases: A generalization of Kruskal rank}

In this section, we show identification without imposing linearity by introducing a newly defined generalized Kruskal rank for integral operators. Our analysis focuses on the effectiveness of this new rank concept, while maintaining other high-level assumptions.

We start with an indicator/sparsity matrix similar to a Jacobian matrix. Let $\mathrm{I}(\cdot)$ be the indicator function, which equals 1 if the statement is true and 0 otherwise.  for  For vectors of  random variables $W \in \mathbb{R}^K$ and $X^* \in \mathbb{R}^L$, we define 
\begin{eqnarray}
    \mathrm{J}(p_{W|X^*})= \left [ \mathrm{I} \left (\frac{\partial \, p_{W_i|X^*}(\cdot|x^*)}{\partial x^*_j} \neq 0 \right) \right ]_{i=1,...,K;j=1,...,L}
\end{eqnarray}
where $\mathrm{J}(p_{W|X^*})$ is a matrix of indicators of whether $X^*_j$ enters the distribution $p_{W_i|X^*}$ for each $W_i$ in vector $W$. Let $W_{1:k}$ be the sub-vector of the first $k$ entries in vector $W$, $I_{K\times K}$ be a $K\times K$ identity matrix, and  $R_{K\times L}$ be a $K \times L$ generic  matrix of unspecified remainders. 

We introduce a generalized Kruskal rank for integral operators, called signal rank, as follows:

\begin{definition}
    For a measurement $X \in \mathbb{R}^{K}$ of a latent variable $X^* \in \mathbb{R}^{L}$, 
    the \textbf{signal rank} of integral operator $L_{X|X^*}$ is defined as the largest integer $\kappa^s= \kappa^s(L_{X|X^*})$ such that:
    For any permutation of $X^*_l$ for $l \in \{1,2,...,L\}$ in $X^*$, there exists a continuous one-to-one mapping $Q:\mathbb{R}^{K} \rightarrow \mathbb{R}^{K} $, which satisfies, for $W=Q(X)$
     \begin{enumerate}
         \item  
         $\mathrm{J}(p_{W|X^*})= \begin{pmatrix}
             I_{\kappa^s \times \kappa^s}  & R_{\kappa^s \times (L-\kappa^s)} \\
             R_{(K-\kappa^s) \times \kappa^s}  & R_{(K-\kappa^s) \times (L-\kappa^s)} \\
         \end{pmatrix}$;
         \item $L_{W_{1:k}|X^*_{1:k}}$ is invertible for all $1 \leq k \leq \kappa^s$.
     \end{enumerate}

\end{definition}
In the previous linear case, the signal rank is equal to the Kruskal rank of the factor loading matrix and a mapping $Q$ can be generated by the Gram–Schmidt process based on the loading matrix. Note that $\mathrm{\kappa^s}(L_{X|X^*}) \in \{0,1,2,...,L\}$. We show that a signal rank condition, similar to a Kruskal rank condition for matrices, is crucial for the identification of distributions of latent variables, while keeping other assumptions at the high level for simplicity. 

We assume
\begin{assumption} \label{assumption nonlinear case}
Probability function $p(X^1,X^2,X^3,X^*)$ is  continuous and satisfies
\begin{enumerate} 
\item $p(X^1,X^2,X^3|X^*)  = p(X^1|X^*) \times  p(X^2|X^*) \times  p(X^3|X^*)$.
\item $p(X^*)=p(X^*_1) \times ...\times p(X^*_L).$ 
\end{enumerate}
\end{assumption}
This assumption leads to 
\begin{eqnarray}
    p(X^1,X^2,X^3) &=&  \int p(X^1|X^*)p(X^2|X^*)p(X^3|X^*)p(X^*_1) \times ...\times p(X^*_L)dX^* 
\end{eqnarray}
Our goal is to identify every probability function within the integral on the right hand side. Given the original permutation of vector $X^*= (X^*_1,...,X^*_{L})^T$, we apply its corresponding mapping to obtain $W^1= Q_1(X^1)$. For $X^2$, we consider a permutation of $X^*$ as follows: 
\begin{equation}
    X^*= (X^*_1 ; X^*_{L-\kappa^s_2+2}, X^*_{L-\kappa^s_2+3},...,X^*_L ; X^*_2,X^*_3,...,X^*_{L-\kappa^s_2+1})^T
\end{equation}
with $W^2= Q_2(X^2)$. We show that a sufficient condition for identification is that    
\begin{equation}
    \kappa^s_1+\kappa^s_2+\kappa^s_3 \geq 2L+2.
\end{equation}
where $\kappa^s_i$ is the signal rank of  $L_{X^i|X^*}$. This condition guarantees that $X^*_1$ is the only comment factor in the first elements in $W^1$ and $W^2$. Therefore, we can show that idiosyncratic elements in $W^1_1$, and $W^2_1$ are integrated out as follows:
\begin{eqnarray} \label{equation W1 W2}
    p(W^1_1,W^2_1) 
    &=& \int p(W^1_1|X^*_1)  p(W^2_1|X^*_1) p(X^*_1) dX^*_1 
\end{eqnarray}  

In order to identify the whole distribution with a relatively simple procedure, we assume
\begin{assumption} \label{assumption full rank for nonlinear}
    $\kappa^s_3=L.$
\end{assumption} 
\noindent 
Kruskal’s theorem does not require Assumption \ref{assumption full rank for nonlinear}, that is, it does not rely on any of the matrices having full rank, yet it is supported by a more than ten-page proof involving highly sophisticated mathematics. The proof becomes far less complicated with one of the three matrices being full rank. That is why we adopt this assumption. It is possible that Assumption \ref{assumption full rank for nonlinear} is not necessary for our results as well, as we show in the linear case before. Given that this section focuses on the effectiveness of the newly defined signal rank, we mimic this relatively simple case in the proof of the Kruskal's theorem.  

Assumption \ref{assumption full rank for nonlinear} implies that $L_{X^3|X^*}$ has a full signal rank with $W^3=Q_3(X^3)$ satisfying \begin{equation}
    p(W^3_{1:L}|X^*)=\Pi_{l=1}^L p(W^3_l|X^*_l).
\end{equation} and that for $i=1,2$
\begin{eqnarray} \label{equation W3 Wi}
   p(W^3_1,W^i_1) 
    &=& \int p(W^3_1|X^*_1)  p(W^i_1|X^*_1) p(X^*_1) dX^*_1 
\end{eqnarray}  
We can show that three joint distributions $p(W^i_1,W^j_1)$ in Equations (\ref{equation W1 W2}) and (\ref{equation W3 Wi}) identifies $L_{W^3_1|X^*_1} D_{X^*_1} (L_{W^3_1|X^*_1})^T $, which contains only one unknown - $p(W^3_1,X^*_1)$. We put this as a high-level normalization assumption as follows: 
\begin{assumption} \label{assumption normalization high level}
There exist $W^3 = Q_3(X^3)$ s.t. 
    $L_{W^3_l|X^*_l} D_{X^*_l} (L_{W^3_l|X^*_l})^T $ uniquely determines $L_{W^3_l|X^*_l}$ and $D_{X^*_l}$, i.e., $p(W^3_1,X^*_1)$, for all $l$.
\end{assumption}
Notice that the operator $L_{W^3_1|X^*_1} D_{X^*_1} (L_{W^3_1|X^*_1})^T $ can be considered as an operator based on the joint distribution of $(W^3_1,\tilde{W}^3_1)$ with 
\begin{eqnarray}
    W^3_1 &=& X^*_1 + e^1 \\
    \tilde{W}^3_1 &=& X^*_1 + \tilde{e}^1 \nonumber
\end{eqnarray}
where $e^1$ and $\tilde{e}^1$ have the same distribution conditional on $X^*_1$. If $X^*_1$, $e^1$, and $\tilde{e}^1$  on the right hand side are jointly independent and have a nonvarnishing ch.f., the Kortlarski's identity implies that the distributions of $(X^*_1, e^1, \tilde{e}^1)$, equivalently $p(W^3_1,X^*_1)$,  are all identified with a closed form. Note that these derivations hold for $L_{W^3_l|X^*_l}$ as well. That means that Assumption \ref{assumption normalization high level} implies that $L_{W^3_l|X^*_l}$ for all $l$ are identified. 

Finally, we show the whole distribution $p(X^1,X^2,X^3,X^*)$ is identified from
\begin{eqnarray}
    p(X^1,X^2,W^3) &=& \int p(W^3|X^*)p(X^1,X^2,X^*)dX^*. \nonumber\\ 
\end{eqnarray}
Because we have identified $L_{W^3_l \mid X^*_l}$ for all $l$ and they are invertible, we can identify \\ \( p(X^2, X^3, X^*) \). Given the mapping $Q^3$ is one-to-one and continuous, we also identify the distribution of $p(X^3 \mid X^*)$ from $p(W^3 \mid X^*)$. That means we have identified 
\begin{equation}
    p(X^1,X^2,X^3,X^*) = p(X^3|X^*) p(X^2,X^3,X^*)
\end{equation}
We summarize the result as follows:

\begin{theorem} \label{theorem identification general}
    Suppose that Assumptions \ref{assumption nonlinear case}, \ref{assumption full rank for nonlinear}, and  \ref{assumption normalization high level}  hold. Then, $p(X^i|X^*)$ for $i=1,2,3$ and  $p(X^*)$  are unique, up to the permutation of $X^*_l$ for $l=1,2,...,L$, in 
     \begin{equation}
          p(X^1,X^2,X^3) = \int p(X^1|X^*)p(X^2|X^*)p(X^3|X^*)p(X^*)dX^*
     \end{equation}
     if 
     \begin{equation} \label{equation: signal rank condition}
\kappa^s(L_{X^1|X^*})+\kappa^s(L_{X^2|X^*})+\kappa^s(L_{X^3|X^*}) \geq 2L+2.
     \end{equation}
     where $\kappa^s(L_{X^i|X^*})$ is the signal rank of $L_{X^i|X^*}$.
\end{theorem}

\textbf{Proof}: See the Appendix.

\section{Summary}

This paper establishes new identification results for multivariate measurement-error models in which all observed measurements contain correlated noise and none provides an injective mapping of the latent vector. By exploiting the structure of third-order cross-moments, I construct a three-way tensor whose decomposition is uniquely determined under Kruskal’s rank condition. This tensor decomposition identifies the factor loading matrices up to permutation and scaling. Building on a linear structure, the paper then identifies the full distribution of each latent component by constructing appropriate measurements and applying either scalar or multivariate versions of Kotlarski’s identity. As a result, the joint distribution of the latent vector and all measurement errors is determined nonparametrically without requiring injective measurements. With the injectivity, we also achieve the identification under testable and less restrictive conditions.

The results in this paper widen the scope of what can be identified in measurement-error models, demonstrating that multivariate latent structures can be recovered in settings that would otherwise appear underidentified using existing techniques. The identification in the linear case is constructive in the sense that it leads to a straightforward estimation procedure. This makes the approach useful for empirical work involving noisy regressors, survey misreporting, and multidimensional factor models in economics and the social sciences.

Furthermore, this paper provides a generalized Kruskal rank - signal rank - for integral operators, and show that identification holds for general nonlinear models under a signal rank condition similar to the well-known Kruskal rank condition. The newly-defined signal rank is consistent with the existing Kruskal rank for matrices and the injectivity of integral operators, and can be used to describe how informative a measurement is in the multivariate case. Although our identification of nonlinear models is not as constructive as that of linear models, the results here can be seen as a small step toward developing a continuous analogue of Kruskal’s classical theorem. As shown in \cite{allman2009identifiability}, Kruskal’s theorem delivers strong identification results for models with discrete latent variables. Extending these ideas to continuous latent variables would require a version of Kruskal’s theorem that works in a continuous setting, which, to the best of my knowledge, is still an open question. The Hu–Schennach theorem provides a partial analogue, but only under an injectivity condition. This paper provides constructive identification for linear measurement error models and defines signal rank of integral operators for the multivariate case, which moves the literature closer towards a solution of the important open question.

\section{Appendix}
\subsection{Proof of Lemma \ref{Lemma loading}}
We start with the linear model 
\begin{eqnarray} 
X^1 &=& M^1 X^* + \varepsilon^1  \\
X^2 &=& M^2 X^* +\varepsilon^2 \nonumber \\
X^3 &=& M^3 X^* +\varepsilon^3  \nonumber
\end{eqnarray}
Assumption \ref{assumption cond mean independence}.1 implies that for $i \neq j$ and $i \neq k$
$$E(\varepsilon^i \times X^j \times X^k)=0.$$
We then have
\begin{eqnarray} 
E (X_i^{1} \times X_u^2 \times X_v^3) 
&=&  \sum_{j=1} \sum_{k=1} \sum_{l=1} E( m^1_{i,j} X^*_j \times m^2_{u,k} X^*_k \times m^3_{v,l} X^*_l )  \nonumber \\
&=&  \sum_{j=1} \sum_{k=1} \sum_{l=1}  m^1_{i,j} \times m^2_{u,k} \times m^3_{v,l} \times   E [X^*_j X^*_k X^*_l]  
\end{eqnarray}
Assumption \ref{assumption cond mean independence}.2 implies that for $l \neq j$ and $l \neq k$
\begin{eqnarray} 
E [X^*_j X^*_k X^*_l]
&=&  E \left(E [X^*_j X^*_k | X^*_l] \times X^*_l \right )  \\
&=&  E \left(E [X^*_j X^*_k ] \times X^*_l \right ) \nonumber \\
&=&  E [X^*_j X^*_k ] E(X^*_l) \nonumber\\
&=& 0 \nonumber
\end{eqnarray}
Therefore, 
\begin{eqnarray} 
E (X_i^{1} \times X_u^2 \times X_v^3) 
&=&  \sum_{l=1}^{l=L} m^1_{i,l} \times m^2_{u,l} \times m^3_{v,l} \times   E [(X^*_l)^3]
\end{eqnarray}
These third order moments form a 3-way tensor. The Kruskal's Theorem (\cite{Kruskal1977}) implies that the elements on the right hand side, i.e., $M^1$, $M^2$, $M^3$ and $E [(X^*_i)^3 ] $, are unique up to permutation and scaling under the Kruskal rank condition  in Equation (\ref{equation: k rank}) . That means that we have identified the factor loadings. QED.

\subsection{Proof of Theorem \ref{theorem identification distr}}

Given the independence assumption 
\begin{equation}
    p(\varepsilon^1,\varepsilon^2,\varepsilon^3) =  p(\varepsilon^1) \times  p(\varepsilon^2) \times  p(\varepsilon^3)
\end{equation}
and 
\begin{equation}
    p(X^*)=p(X^*_1) \times ...\times p(X^*_L)
\end{equation}
we first show how to identify the distribution of each $X^*_l$. Then, the distributions of $\varepsilon^i$  are identified by deconvolution from $X^i$. 

To identify the distribution of $X^*_1$, we find two measurements of $X^*_l$ satisfying the conditions to use the Kotlarski's identity. Let $\kappa^i = \kappa(M^i)$. That means one can generate a $\kappa^1 \times \kappa^1$ identity matrix from $M^i$ by applying the Gram-Schmidt algorithm to the rows. Therefore, there exist $Q^1$ and $Q^2$ s.t.
\begin{eqnarray} 
W^1 \equiv Q^1X^1 
&=&  Q^1M^1X^* + Q^1\varepsilon^{1} \\
&=& \begin{pmatrix}
I_{\kappa^1 \times \kappa^1} & R^1_{\kappa^1\times (L-\kappa^1)} \\
0  &R^1_{(K^1-\kappa^1)\times (L-\kappa^1)}
\end{pmatrix}
X^*  + Q^1\varepsilon^{1} \nonumber \\
&\equiv& \begin{pmatrix}
1 & 0 & R^1_{1\times (L-\kappa^1)} \\
0& I_{(\kappa^1-1) \times (\kappa^1-1)} & R^1_{(\kappa^1-1)\times (L-\kappa^1)} \\
0  & 0& R^1_{(K^1-\kappa^1)\times (L-\kappa^1)}
\end{pmatrix}
X^*  + Q^1\varepsilon^{1} \nonumber
\end{eqnarray}
where $R^1$ refers to a generic remainder matrix in $M^1$ with its dimensions in the subscript. Matrices $R^1_{(K^1-\kappa^1)\times (L-\kappa^1)}$  may not appear if $\kappa^1=K^1$. The first element in $W^1$ can be considered as a measurement of $X^*_1$ with other elements as its measurement error. Let $[M]_1$ stand for the first row of matrix $M$. WLOG, we consider the first element in $X^*$
\begin{eqnarray} 
W^1_1 
&=&  X^*_1 +  R^1_{1\times (L-\kappa^1)} (X^*_{\kappa^1+1},...,X^*_L)^T+  [Q^1]_1\varepsilon^{1} \nonumber \\
 &\equiv& X^*_1 + e_1 \label{equation e1}
\end{eqnarray}
Next, we  generate a second measurement of $X^*_1$ from $X^2$, in which there is no other common factor than $X^*_1$. We will show that the Kruskal rank condition guarantees the existence of such a measurement. For that reason, we generate a $\kappa^2 \times \kappa^2$ identity matrix from the right side, i.e., starting from $X^*_L$ in the matrix $M^2$. 

Similarly, there exist $Q^2$ corresponding to $M^2$ in measurement $X^2$ s.t.
\begin{eqnarray} 
W^2 \equiv Q^2X^2 
&=&  Q^2M^2 X^* + Q^2\varepsilon^{2}  \\
&=& \begin{pmatrix}
R^2_{(K^2-\kappa^2)\times (L-\kappa^2)} & 0\\
R^2_{(\kappa^2)\times (L-\kappa^2)} & I_{\kappa^2 \times \kappa^2}
\end{pmatrix}X^* + Q^2\varepsilon^{2} \nonumber \\
&\equiv& \begin{pmatrix}
R^2_{(K^2-\kappa^2)\times (L-\kappa^2)} & 0 &0 \\
R^2_{1\times (L-\kappa^2)}  & 1 & 0 \\
R^2_{(\kappa^2-1)\times (L-\kappa^2)} & 0   & I_{(\kappa^2-1) \times (\kappa^2-1)}
\end{pmatrix}X^* + Q^2\varepsilon^{2} \nonumber \\
&\equiv& \begin{pmatrix}
\begin{pmatrix}
R^2_{(K^2-\kappa^2)\times (L-\kappa^2)} & 0 \\
R^2_{1\times (L-\kappa^2)}  & 1
\end{pmatrix}& \begin{pmatrix}
    0\\0
\end{pmatrix} \\
\begin{pmatrix} R^2_{(\kappa^2-1)\times (L-\kappa^2)} & 0  \end{pmatrix} & I_{(\kappa^2-1) \times (\kappa^2-1)}
\end{pmatrix}X^* + Q^2\varepsilon^{2} \nonumber \\
&\equiv& \begin{pmatrix}
R^2_{(K^2-\kappa^2+1)\times (L-\kappa^2+1)} & 0 \\
R^2_{(\kappa^2-1)\times (L-\kappa^2+1)}  & I_{(\kappa^2-1) \times (\kappa^2-1)}
\end{pmatrix}X^* + Q^2\varepsilon^{2} \nonumber
\end{eqnarray}
where $R^2$ refers to a generic remainder matrix in $M^2$ with its dimensions in the subscript. In these derivations, we only reorganize the matrix so that we are able to locate a suitable measurement. We then pick an element in $W^2$ which does not contain $(X^*_{\kappa^1+1},...,X^*_L)$, which are in $W^1_1$, but does contain $X^*_1$.  We consider the i-th row of $R^2_{(K^2-\kappa^2+1)\times (L-\kappa^2+1)}$ for all $i\leq L-\kappa^2+1$. There must be a nonzero entry $[R^2_{(K^2-\kappa^2+1)\times (L-\kappa^2+1)}]_{i,1} \neq 0$ in the first column (corresponding to $X^*_1$) in $R^2_{(K^2-\kappa^2+1)\times (L-\kappa^2+1)}$. Otherwise, the Kruskal rank has to be smaller than $\kappa^2$. We then consider
\begin{eqnarray} 
W^2_{i} 
&=&  [R^2_{(K^2-\kappa^2+1)\times (L-\kappa^2+1)}]_{i,1}X^*_1 + \sum_{j=2}^{L-\kappa^2+1} [R^2_{(K^2-\kappa^2+1)\times (L-\kappa^2+1)}]_{i,j}X^*_j +  [Q^2]_i\varepsilon^{2} \nonumber \\
 &\equiv& r X^*_1 + e_2 \label{equation e2}
\end{eqnarray}
In summary, $e_2$ in Equation (\ref{equation e2}) contains  $(X^*_{2},...,X^*_{L-\kappa^2+1}  ,\varepsilon^{2})$ and $e_1$ in Equation (\ref{equation e1}) contains $(X^*_{\kappa^1+1},...,X^*_L  ,\varepsilon^{1})$. 
Notice that the Kruskal rank condition in Equation (\ref{equation: k rank}) guarantees that 
\begin{equation}
    \kappa^1+\kappa^2 \geq L+2
\end{equation}
because $L \geq \kappa^3 $ and $\kappa^1+\kappa^2+\kappa^3 \geq 2L+2$. That means
\begin{equation}
    \kappa^1+1 > L-\kappa^2+1.
\end{equation}
Therefore, $(X^*_1, e^1, e^2)$ have no common elements and therefore are jointly independent.

We then reveal the distribution of $X^*_1$ from two scalar measurements $(W^1_1,W^2_1)$ given that $r\neq0$ has been identified. For each latent dimension, $p(X^*_1)$ is identified from its ch.f. by Kotlarski’s Identity
\begin{equation}
    \phi _{X^*_1}\left( t\right) =\exp \left( \int_{0}^{t}\frac{\left[ \frac{%
\partial }{\partial t_{2}}\phi _{W^1_1,W^2_1}\left( s,\frac{t_{2}}{\gamma}\right) \right]
_{t_{2}=0}}{\phi _{W^1_1}\left( s\right) }ds\right) . 
\end{equation}
where  $\phi_X$ is ch.f. of $X$. 
Given the definition of the Kruskal rank, this identification procedure applies to all $X^*_l$ with any permutation of the columns in $M^i$ or $X^*_l$ in $X^*$. That means distributions of $X^*$, $\varepsilon^1$, $\varepsilon^2$ are nonparametrically identified with a closed-form solution as follows:
\begin{eqnarray} 
\phi _{X^*} &= &\phi _{X^*_1}\times ...\times\phi _{X^*_L} \\
\phi_{\varepsilon^i} &=& \frac{\phi_{X^i}}{\phi_{M^i X^*} }
\end{eqnarray}
where $\phi_{X^i}=\phi_{\varepsilon^i}\phi_{M^i X^*}   \neq 0$. 
In summary, we have identified the distributions of $X^*_1$, ... ,$X^*_l$, $\varepsilon^1$, $\varepsilon^2$, $\varepsilon^3$ from the joint distribution $p(X^1,X^2,X^3)$. In other words, the joint distribution $p(X^1,X^2,X^3)$ uniquely determines the joint distribution $p(X^1,X^2,X^3,X^*)$, where
\begin{equation}
    p(X^1,X^2,X^3,X^*) = p_{\varepsilon^1}(X^1-M^1 X^*)p_{\varepsilon^2}(X^2-M^2 X^*)p_{\varepsilon^3}(X^3-M^3 X^*)p(X^*_1)...p(X^*_L).
\end{equation}
QED.

\subsection{Proof of Proposition \ref{theorem identification correlated}}
 First, Lemma \ref{Lemma loading} suggests that Assumptions \ref{assumption cond mean independence} and the Kruskal rank condition in Equation (\ref{equation: k rank})  guarantee that factor loading matrices $M^1$, $M^2$, $M^3$ in Equation \ref{equation model} are identified up to permutation and scaling.
 We then show that all the distributions are still identified when the elements in $X^*$ are correlated. Assumption \ref{assumption identification correlated X^*}.2 implies that we can use $X^1$ a primary measurement of $X^*$ and combine $X^2$ and $X^3$ as a secondary measurement. We then use the multivariate Kotlarski's identity to identify the joint distribution of $X^*$. 

The full column rank conditions in Equation (\ref{equation full rank}) guarantees that there exist $Q_1$ and $Q_{23}$ such that 
\begin{eqnarray} 
W^1 \equiv Q^1X^1  
&=& X^*  + Q^1\varepsilon^{1}  \\
W^{23} \equiv Q^{23} \begin{pmatrix}
    X^2 \\ X^3
\end{pmatrix} &=& X^*  + Q^{23} \begin{pmatrix}
    \varepsilon^{2} \\ \varepsilon^{3}
\end{pmatrix}  
\end{eqnarray}
We can then identify the distribution $p(X^*)$ from its ch.f. as follows:
for $t,s \in \mathbb{R}^L$ and $\lambda \in [0,1]$
\begin{equation}
    \phi _{X^*}\left( t\right) =\exp \left( \int_{0}^{1}\frac{\left[ \nabla_{s} \phi _{W^1,W^{23}}\left( \lambda t,s\right) \right]^T
_{s=0}t}{\phi _{W^1}\left(\lambda t \right) }d \lambda \right) . 
\end{equation}
The distributions of all other elements are also identified. We have 
\begin{eqnarray} 
\phi_{\varepsilon^i} &=& \frac{\phi_{X^i}}{\phi_{M^i X^*} }
\end{eqnarray}
where $\phi_{X^1}=\phi_{\varepsilon^i}\phi_{M^i X^*}   \neq 0$ implies that $\phi_{M^i X^*} \neq 0$
In summary, we have identified the distributions of $X^*$, $\varepsilon^1$, $\varepsilon^2$, $\varepsilon^3$ from the joint distribution $p(X^1,X^2,X^3)$. In other words, the joint distribution $p(X^1,X^2,X^3)$ uniquely determines the joint distribution $p(X^1,X^2,X^3,X^*)$, where
\begin{equation}
    p(X^1,X^2,X^3,X^*) = p_{\varepsilon^1}(X^1-M^1 X^*)p_{\varepsilon^2}(X^2-M^2 X^*)p_{\varepsilon^3}(X^3-M^3 X^*)p(X^*).
\end{equation}
QED.

\subsection{Proof of Theorem \ref{theorem identification general}}
Here we show the identification for the genearl nonlinear case. Assumption \ref{assumption nonlinear case} leads to 
\begin{eqnarray}
    p(X^1,X^2,X^3) &=& \int p(X^1|X^*)p(X^2|X^*)p(X^3|X^*)p(X^*)dX^* \nonumber \\
    &=& \int p(X^1|X^*)p(X^2|X^*)p(X^3|X^*)p(X^*_1) \times ...\times p(X^*_L)dX^* 
\end{eqnarray}
The goal is to identify every probability function within the integral on the right hand side. Given the original permutation of vector $X^*= (X^*_1,...,X^*_{L})^T$, we apply its corresponding mapping to obtain $W^1= Q_1(X^1)$. For $X^2$, we consider a permutation of $X^*$ as follows: 
\begin{equation}
    X^*= (X^*_1 ; X^*_{L-\kappa^s_2+2}, X^*_{L-\kappa^s_2+3},...,X^*_L ; X^*_2,X^*_3,...,X^*_{L-\kappa^s_2+1})^T
\end{equation}
with $W^2= Q_2(X^2)$. Let 
\begin{equation}
    \kappa^s_i := \kappa^s(L_{X^i|X^*}).
\end{equation}

We start with the joint distribution
\begin{eqnarray}
    p(W^1,W^2) &=& \int p(W^1|X^*)p(W^2|X^*)p(X^*_1) \times ...\times p(X^*_L)dX^* 
\end{eqnarray}    
For the first elements in $W^1$ and $W^2$ satisfy
\begin{eqnarray}
   && p(W^1_1,W^2_1) \\
   &=& \int p(W^1_1|X^*)p(W^2_1|X^*)p(X^*_1)  ... p(X^*_L)dX^*  \nonumber \\
     &=& \int p(W^1_1|X^*_1,...,X^*_{L})  p(W^2|X^*_1 ; X^*_{L-\kappa^s_2+2}, X^*_{L-\kappa^s_2+3},...,X^*_L ; X^*_2,X^*_3,...,X^*_{L-\kappa^s_2+1})  \nonumber  \\
     &&\times p(X^*_1) \times ...\times p(X^*_L)dX^*   \nonumber 
\end{eqnarray}  
Given 
\begin{equation}
    \mathrm{J}(p_{W^1|X^*}) = \begin{pmatrix}
    I_{\kappa^s_1 \times \kappa^s_1}  & R_{\kappa^s_1 \times (L-\kappa^s_1)} \\
    R_{(K_1-\kappa^s_1) \times \kappa^s_1}  & R_{(K_1-\kappa^s_1) \times (L-\kappa^s_1)} 
         \end{pmatrix}
\end{equation}
and 
\begin{equation}
    \mathrm{J}(p_{W^2|X^*}) =\begin{pmatrix}
             I_{\kappa^s_2 \times \kappa^s_2}  & R_{\kappa^s_2 \times (L-\kappa^s_2)}\\
            R_{(K_2-\kappa^s_2) \times \kappa^s_2}  & R_{(K_2-\kappa^s_2) \times (L-\kappa^s_2)} 
         \end{pmatrix}
\end{equation}
they imply that $W^1_1$ does not contain $(X^*_2,...,X^*_{\kappa^s_1})$ and $W^2_1$ does not contain $(X^*_{L-\kappa^s_2+2},...,X^*_L)$ so that 
\begin{eqnarray}
    && p(W^1_1,W^2_1) \\
    &=& \int p(W^1_1|X^*)p(W^2_1|X^*)p(X^*_1)  ...p(X^*_L)dX^*  \nonumber  \\
     &=& \int p(W^1_1|X^*_1,X^*_{\kappa^s_1+1},...,X^*_{L}) p(W^2|X^*_1,X^*_2,...,X^*_{L-\kappa^s_2+1})  p(X^*_1) ... p(X^*_L)dX^*   \nonumber 
\end{eqnarray}  
Note that $\mathrm{\kappa^s}_3 \leq L$. The signal rank condition in Equation \ref{equation: signal rank condition} implies  
    \begin{equation}
        \kappa^s_1+\kappa^s_2 \geq L+2
    \end{equation}
     and
     \begin{equation}
         \kappa^s_1+1  > L-\kappa^s_2 +1 
     \end{equation}
which implies that $X^*_1$ is the only common element between $W^1_1$, and $W^2_1$. Therefore, idiosyncratic elements in $W^1_1$, and $W^2_1$ are integrated out as follows:
\begin{eqnarray}
  &&  p(W^1_1,W^2_1) \\
     &=& \int p(W^1_1|X^*_1,X^*_{\kappa^s_1+1},...,X^*_{L}) \times \nonumber \\
     &&\times p(W^2_1|X^*_1,X^*_2,...,X^*_{L-\kappa^s_2+1})  p(X^*_1)  ... p(X^*_{L})dX^* \nonumber \\
    &=& \int p(W^1_1,X^*_1,X^*_{\kappa^s_1+1},...,X^*_{L}) \times \nonumber \\
     &&\times p(W^2_1|X^*_1,X^*_2,...,X^*_{L-\kappa^s_2+1}) p(X^*_2)  ... p(X^*_{L-\kappa^s_2+1}) dX^* \nonumber \\
    &=& \int p(W^1_1,X^*_1)  p(W^2_1|X^*_1,X^*_2,...,X^*_{L-\kappa^s_2+1})  p(X^*_2)  ... p(X^*_{L-\kappa^s_2+1})dX^*_1 dX^*_2 ...d X^*_{L-\kappa^s_2+1} \nonumber \\ 
    &=& \int p(W^1_1|X^*_1)  p(W^2_1,X^*_1,X^*_2,...,X^*_{L-\kappa^s_2+1})  dX^*_1 dX^*_2 ...d X^*_{L-\kappa^s_2+1}\nonumber \\ 
    &=& \int p(W^1_1|X^*_1)  p(W^2_1|X^*_1) p(X^*_1) dX^*_1 \nonumber
\end{eqnarray}  

Next, Assumption \ref{assumption full rank for nonlinear} implies that $L_{X^3|X^*}$ has a full signal rank with $W^3=Q_3(X^3)$ and 
\begin{equation}
    \mathrm{J}(p_{W^3|X^*}) = \begin{pmatrix}
             I_{L \times L}  \\
            R_{(K_3-L) \times L}  
         \end{pmatrix}
\end{equation}
and therefore,
\begin{equation}
    p(W^3_{1:L}|X^*)=\Pi_{l=1}^L p(W^3_l|X^*_l).
\end{equation}
We then consider
\begin{eqnarray}
  &&  p(W^1_1,W^3_1) \\
     &=& \int p(W^1_1|X^*_1,X^*_{\kappa^s_1+1},...,X^*_{L})  p(W^3_1|X^*_1) p(X^*_1)  ... p(X^*_{L})dX^*  \nonumber \\
    &=& \int p(W^1_1|X^*_1)  p(W^3_1|X^*_1) p(X^*_1) dX^*_1 \nonumber
\end{eqnarray}  
Similarly, we have
\begin{eqnarray}
   p(W^3_1,W^2_1)  &=& \int p(W^3_1|X^*_1)   p(W^2_1|X^*_1,X^*_2,...,X^*_{L-\kappa^s_2+1})  p(X^*_1)  ... p(X^*_{L})dX^*  \\
   &=& \int p(W^3_1|X^*_1)  p(W^2_1|X^*_1) p(X^*_1) dX^*_1  \nonumber
\end{eqnarray}  
In operator form, we have
\begin{eqnarray}
L_{W^1_1,W^2_1} &=& L_{W^1_1|X^*_1}D_{X^*_1}(L_{W^2_1|X^*_1})^T   \\
L_{W^1_1,W^3_1} &=& L_{W^1_1|X^*_1}D_{X^*_1}(L_{W^3_1|X^*_1})^T  \nonumber \\
L_{W^3_1,W^2_1} &=& L_{W^3_1|X^*_1}D_{X^*_1}(L_{W^2_1|X^*_1})^T \nonumber
\end{eqnarray}  
Given that these operators are invertable, we have
\begin{eqnarray}
  && L_{W^3_1,W^2_1} (L_{W^1_1,W^2_1})^{-1} L_{W^1_1,W^3_1}  \nonumber \\
   &=& L_{W^3_1|X^*_1}D_{X^*_1}(L_{W^2_1|X^*_1})^T ( L_{W^1_1|X^*_1}D_{X^*_1}(L_{W^2_1|X^*_1})^T)^{-1} L_{W^1_1|X^*_1}D_{X^*_1}(L_{W^3_1|X^*_1})^T  \nonumber \\
   &=& L_{W^3_1|X^*_1} D_{X^*_1} (L_{W^3_1|X^*_1})^T 
\end{eqnarray}
where the left hand side is identified from three observed joint distributions. Assumption \ref{assumption normalization high level} then implies that  $L_{W^3_l|X^*_l}$ and $D_{X^*_l}$, i.e., $p(W^3_1,X^*_1)$, for all $l$ are identified.

Finally, we show the whole distribution $p(X^1,X^2,X^3,X^*)$ is identified. We consider
\begin{eqnarray}
    p(X^1,X^2,W^3) &=& \int p(W^3|X^*)p(X^1,X^2,X^*)dX^* \nonumber \\
    &=& \int p(W^3_1|X^*_1)... p(W^3_L|X^*_L)p(X^1,X^2,X^*)dX^* 
\end{eqnarray}
Because we have identified $L_{W^3_l|X^*_l}$ for all  $l$ and they are invertible, we can identify\\ $p(X^2,X^3,X^*)$. Given the mapping $Q^3$ is one-to-one and continuous, we also identify the distribution of $p(X^3|X^*)$ from $p(W^3|X^*)$. That means we have identified 
\begin{equation}
    p(X^1,X^2,X^3,X^*) = p(X^3|X^*) p(X^2,X^3,X^*).
\end{equation}

QED.

\bibliographystyle{aer}
\bibliography{Ref-all}

@article{Hu2008discrete,
  author       = {Hu, Yingyao},
  title        = {Identification of Nonparametric Measurement Error Models with Discrete Data},
  journal      = {Econometrica},
  volume       = {76},
  number       = {1},
  pages        = {195--216},
  year         = {2008}
}

@article{kotlarski1965pairs,
  title={On Pairs of Independent Random Variables Whose Product Follows the Gamma Distribution},
  author = {Kotlarski, Ignacy},
  journal={Biometrika},
  pages={289--294},
  year={1965},
  publisher={JSTOR}
}

@article{Kruskal1977,
  author       = {Kruskal, Joseph B.},
  title        = {Three-way arrays: Rank and uniqueness of trilinear decompositions, with application to arithmetic complexity and statistics},
  journal      = {Linear Algebra and Its Applications},
  volume       = {18},
  number       = {2},
  pages        = {95--138},
  year         = {1977}
}

@article{CarrollChang1970,
  author       = {Carroll, J. Douglas and Chang, Jih-Jie},
  title        = {Analysis of Individual Differences in Multidimensional Scaling via an N-Way Generalization of ``Eckart-Young'' Decomposition},
  journal      = {Psychometrika},
  volume       = {35},
  number       = {3},
  pages        = {283--319},
  year         = {1970}
}

@article{Harshman1970,
  author       = {Harshman, Richard A.},
  title        = {Foundations of the PARAFAC procedure: Models and conditions for an ``explanatory'' multimodal factor analysis},
  journal      = {UCLA Working Papers in Phonetics},
  volume       = {1},
  number       = {16},
  pages        = {1--84},
  year         = {1970}
}

@article{Sidiropoulos2000,
  author       = {Sidiropoulos, Nicholas D. and Bro, Rasmus and Giannakis, Georgios B.},
  title        = {Parallel Factor Analysis in Sensor Array Processing},
  journal      = {IEEE Transactions on Signal Processing},
  volume       = {48},
  number       = {8},
  pages        = {2377--2388},
  year         = {2000}
}

@article{BonhommeJochmansRobin2016,
  author       = {Bonhomme, St{\'e}phane and Jochmans, Koen and Robin, Jean-Marc},
  title        = {Nonparametric identification in finite mixtures of nonparametric product measures},
  journal      = {Journal of the Royal Statistical Society: Series B},
  volume       = {78},
  number       = {4},
  pages        = {1--32},
  year         = {2016}
}

@article{hu_schennach,
	Author = {Hu, Yingyao and Schennach, Susanne},
	Journal = {Econometrica},
	Pages = {195--216},
	Title = {Instrumental Variable Treatment of Nonclassical Measurement Error Models},
	Volume = {76},
	Year = {2008}}

@article{allman2009identifiability,
  title={Identifiability of Parameters in Latent Structure Models with Many Observed Variables},
  author={Allman, Elizabeth S and Matias, Catherine and Rhodes, John A},
  journal={The Annals of Statistics},
  pages={3099--3132},
  year={2009},
  publisher={JSTOR}
}

@article{HuShiu2022,
  title={{A Simple Test of Completeness in a Class of Nonparametric Specification}},
  author={Hu, Yingyao and Shiu, Ji-Liang},
  journal={Econometric Reviews},
  volume={41},
  number={4},
  pages={373--399},
  year={2022},
  publisher={Taylor \& Francis}
}
\end{document}